\documentstyle[twocolumn,prb,aps, epsf]{revtex}

\begin{document}
\draft
\title{Unified theory of phase separation and charge ordering
in doped manganite perovskites }
\author{Shun-Qing Shen and Z. D. Wang}
\address{Department of Physics,The University of Hong Kong, Pokfulam, Hong Kong, China}
\date{\today}

\twocolumn[\hsize\textwidth\columnwidth\hsize\csname@twocolumnfalse\endcsname

\maketitle

\begin{abstract}
A unified theory is developed to explain various types of electronic
collective behaviors in doped manganites R$_{1-x}$X$_x$MnO$_3$ (R = La,
Pr,Nd etc. and X = Ca, Sr, Ba etc.). Starting from a realistic electronic
model, we derive an effective Hamiltonianis by ultilizing the projection
perturbation techniques and develop a spin-charge-orbital coherent state
theory, in which the Jahn-Teller effect and the orbital degeneracy of e$_g$
electrons in Mn ions are taken into account. Physically, the experimentally
observed charge ordering state and electronic phase separation are two
macroscopic quantum phenomena with opposite physical mechanisms, and their
physical origins are elucidated in this theory. Interplay of the Jahn-Teller
effect, the lattice distortion as well as the double exchange mechanism
leads to different magnetic structures and to different charge ordering
patterns and phase separation.
\end{abstract}

]

\section{Introduction}

Charge ordering states (CO) \cite
{Jirak85,Knizek92,Kuwahara95,Radaelli95,Ramirez96,Chen96,Tokura96} and
electronic phase separation \cite{Allodi97,Hennion98} (PS) are two of the
macroscopic quantum phenomena observed experimentally in doped manganites R$%
_{1-x}$X$_{x}$MnO$_{3}$ (R=La, Pr, Nd etc. and X=Ca, Sr, Ba etc.). This
seems puzzling since these two phenomena have completely opposite physical
mechanisms. CO near $x=0.5$ is a regular alignment of Mn$^{3+}$ and Mn$^{4+}$
in the real space. It is well known that the Wigner lattice is expected to
be stabilized when the repulsive potential between charge carriers dominates
over the kinetic energy of the carriers. In this respect CO is expected to
form in manganites due to strong repulsion between charge carriers.
Oppositely, PS in doped manganites near $x=0$ is characteristic of two
regions of rich- and poor-density of charge carriers with ferromagnetic (FM)
and antiferromagnetic (AF) correlations, respectively. A uniform-density
state is unstable when the charge carriers are subjected to a strong
attractive interaction, as discussed in high T$_{c}$ superconductors.\cite
{Kivelson94} It should be an strong attraction which drives the charge
carriers to the electronic PS. On the other hand, various types of magnetic
structures and orbital ordering states were also observed experimentally. CO
and PS are definitely associated with these structures. For instance, CO
with the $(\pi ,\pi ,0)$ patterns occurs under the C-type antiferromagnetic
(AF) background, and PS near $x=0$ occurs under the A-type AF background.
The field-induced melting effect of CO shows that the CO\ decreases and
eventually disappears while an external magnetic field increases.\cite
{Tomioka95} The field destroys the AF correlation, and the disappearance of
CO and AF indicates their close relations and the possible relation between
AF\ correlation and repulsive interaction of charge carriers. Hence
experimental observations of CO and PS in manganites strongly suggest that
the sign of effective interaction between the charge carriers should depend
on the dopant concentrations and the magnetic structures.

There has been considerable theoretical work motivated by the experimental
research on manganese oxides. Most theoretical efforts focus on
understanding metallic ferromagnetism and its connection to unusual
transport properties\cite
{Zener51,Anderson60,Kubo72,Millis95,Sheng97,Alexandrov99}. The scenario of
double exchange mechanism is extensively accepted to explain the metallic
ferromagnetism. However, we still lack a comprehensive picture for the
physical origins of the PS and CO in doped manganites. In a simplified
one-band model, PS was studied analytically and numerically. \cite
{Yunoki97,Shen98,Arovas98} An attraction between the charge carriers caused
by the superexchange coupling is responsible for the instability of a
uniform-density state.\cite{Shen98} In the vicinity of $x=0.5$, the
mechanisms of both long-range Coulomb interaction and the particle-hole
interaction for the CO were proposed.\cite{Tomioka95,Lee97,Shen99b} However,
PS and CO cannot be explained simultaneously in the same one-band model as
strong long-range Coulomb interaction does not favor forming the PS near $x=0
$. Other properties, such as the ferromagnetism at $x=0$ and anomaly optical
conductivity, also suggest that the double degeneracy of e$_{g}$ orbital,
which is neglected in the one-band model, should be included. The phase
diagrams of doped manganites, especially for magnetic and orbital ordering,
have been investigated.\cite
{Kugel73,Inoue95,Ishihara97,Shiba97,Brink99,Yunoki98} Recent overviews for
doped manganites are seen in Ref.\cite{Ramirez97,Moreo99}.

In this paper, we explore the origins of PS and CO in doped manganites and
establish a unified theory for these two phenomena. The paper is organized
as follows. An effective Hamiltonian is derived in Section II. Starting from
a realistic electronic Hamiltonian with strong electron correlations,
several virtual processes of superexchange are considered and an effective
Hamiltonian is derived by means of the projective perturbation approach and
Schwinger boson formalism. A theory of spin-charge-orbital coherent state is
presented in Section III. Close connections of the PS and the CO with
various types of AF are elucidated in Section IV. We also show that the
Jahn-Teller (JT) effect and lattice distortion play important roles in
stabilizing the magnetic structures. Some discussions and a brief summary
are given in Section V. Detailed derivation of the effective Hamiltonian up
to the second order is presented in Appendix A, and the three-site
interaction terms are given in Appendix B.

\section{Effective Hamiltonian: a projection perturbation approach}

\subsection{Model Hamiltonian}

Doped manganese oxides, R$_{1-x}$X$_{x}$MnO$_{3}$, can be regarded as a
mixture of Mn$^{3+}$ (3d$^{4}$) and Mn$^{4+}$ (3d$^{3}$) ions. The three
electrons in the outer shell of Mn$^{4+}$ are almost localized in $t_{2g}$
orbit to form spin maximal state with $S=3/2$. In the manganese ion, Mn$%
^{3+} $, apart from the three localizes electrons in $t_{2g}$ orbit, the
fourth $d$ electron locates at $e_{g}$ orbit which is doubly degenerated. $%
e_{g}$ electrons can become delocalized with increasing $x$. From the above
reasoning, an electronic Hamiltonian with orbital degeneracy is put forward
to describe the dominant electron-electron interaction of the system

\begin{eqnarray}
H_{e} &=&\sum_{ij,\gamma ,\gamma ^{\prime },\sigma }t_{ij}^{\gamma \gamma
^{\prime }}c_{i,\gamma ,\sigma }^{\dagger }c_{j,\gamma ^{\prime },\sigma
}-\sum_{i,\gamma }J_{H}{\bf S}_{i}\cdot {\bf S}_{i,\gamma }  \nonumber \\
&&+\sum_{i,\gamma ,\gamma ^{\prime },\sigma ,\sigma ^{\prime }}(1-\delta
_{\gamma ,\gamma ^{\prime }}\delta _{\sigma ,\sigma ^{\prime }})U_{\gamma
\gamma ^{\prime }}{\bf N}_{i,\gamma ,\sigma }{\bf N}_{i,\gamma ^{\prime
},\sigma ^{\prime }}  \nonumber \\
&&-\sum_{i,\gamma \neq \gamma ^{\prime },\sigma ,\sigma ^{\prime
}}J(c_{i,\gamma ,\sigma }^{\dagger }c_{i,\gamma ,\sigma ^{\prime
}}c_{i,\gamma ^{\prime },\sigma ^{\prime }}^{\dagger }c_{j,\gamma ^{\prime
},\sigma }  \nonumber \\
&&+c_{i,\gamma ,\sigma }^{\dagger }c_{i,\gamma ^{\prime },\sigma
}c_{i,\gamma ,\sigma ^{\prime }}^{\dagger }c_{i,\gamma ^{\prime },\sigma
^{\prime }}),  \label{ham}
\end{eqnarray}
where $c_{i,\gamma ,\sigma }^{\dagger }$ and $c_{j,\gamma ,\sigma }$ \ are
the creation and annihilation operators of e$_{g}$ electron at the orbit $%
\gamma $ ($=z$ or $\bar{z}$ where 
and $\left| z\right\rangle \propto (3z^{2}-r^{2})/\sqrt{3}$ and
$\left| \bar{z}\right\rangle \propto x^{2}-y^{2}$,
respectively.) \ of site $i$ with spin $\sigma $ $(=\uparrow $, $\downarrow
) $, respectively. ${\bf N}_{i,\gamma ,\sigma }=c_{i,\gamma ,\sigma
}^{\dagger }c_{i,\gamma ,\sigma }$. ${\bf S}_{i,\gamma }$ is the spin
operator of e$_{g} $ electron and ${\bf S}_{i}$ is the total maximal spin
operator of the three t$_{2g}$ electrons. Here the transfer integrals in the
model are assumed to take a Slater-Koster form given by the hybridization
between e$_{g}$ orbit and nearest oxygen $p$ orbit, and the model has been
extensively studied to understand physics of doped manganite provskites.\cite
{Kugel73,Ishihara97,Shiba97,Brink99} It was already realized that the Hund's
rule coupling, J$_{H}$, between e$_{g}$ electron and three t$_{2g}$
electrons is very strong, which is the main origin of metallic
ferromagnetism in the range of $0.1<x<0.5$. \cite{Zener51}$\ $Most e$_{g}$
electrons favor forming $S=2$ spins with localized t$_{2g}$ electrons in Mn$%
^{3+}$. Usually only the strong Hund's rule coupling J$_{H}$ is taken into
account. However, the on-site Coulomb interaction U (intra orbit $\gamma
=\gamma ^{\prime }$) and U' (inter orbits $\gamma \neq \gamma ^{\prime }$)
are also dominant energy scales and usually larger than J$_{H}$S. We will
see that the strong on-site correlation play an essential role of the
electronic collective behaviors.

\begin{figure}[tbp]
\epsfxsize=8.5cm
\epsfbox{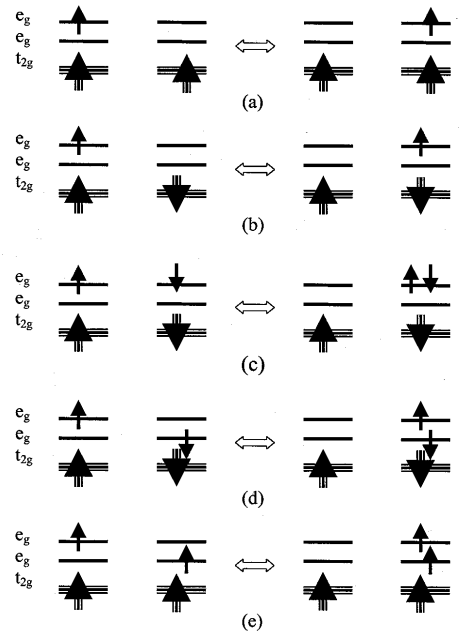}
\caption{Five processes that generate the effective Hamiltonian along the
z-axis. (a) a direct hoping of e$_{g}$ electron from one site to its nearest
neighbour site. The spin of electron must be parallel to the localised spin
at the same site. The (b-e) are four virtual ones to the superexchange
couplings: (b) the mediate state is a single occupancy with spin $S-1/2$;
(c) the mediate state is a double occupancy on the same orbit; (d).the
mediate state is a double occupancy with spin $S+1$ on different orbits; and
(e) the mediate state is a double occupancy with spin $S$ on different
orbits. These five processes corresponds to the five terms in Eq.(2),
respectively. The interactions along other x- and y-axises are derived by
using the symmetry of rotation. }
\end{figure}

Besides the part Hamiltonian of the conduction electrons, we need take into
account other parts of interaction which are believed to affect the phase
diagram of the doped manganites. First, a tiny hopping between t$_{2g}$
electrons produces a weak superexchange AF coupling H$_{AF}$: 
\[
H_{AF}=J_{AF}\sum_{ij}\left( {\bf S}_{i}\cdot {\bf S}_{j}-S^{2}\right) .
\]
Apart from the electronic part of interactions, the JT effect leads to a
static distortion and mixes e$_{g}$ orbits.\cite{Halperin71} According to
Kanamori,\cite{Kanamori61} at x=0, the primary lattice distortion is a
staggered ($\pi ,\pi ,\pi )$ tetragonal distortion of the oxygen octahedra
surrounding the Mn sites, driven by a Jahn-Teller splitting of the outer Mn
d levels. Kanamori's deduction was subsequently confirmed by more detailed
studies of the structure.\cite{Ellemans71} Hence an effective Hamiltonian is
introduced phenomenologically for the JT effect, 
\[
H_{JT}=k\sum_{i,\sigma ,j,\sigma ^{\prime }}({\bf N}_{i,\alpha ,\sigma }-%
{\bf N}_{i,\bar{\alpha},\sigma })({\bf N}_{j,\alpha ,\sigma ^{\prime }}-{\bf %
N}_{j,\bar{\alpha},\sigma ^{\prime }}).
\]
where \ $\alpha $ depends on the direction of ${\bf r}_{i}-{\bf r}_{j}.$ The
tetragonal crystal field H$_{z}$ will lead to the anisotropic magnetic
structure of the system:\cite{Wollan55} 
\[
H_{z}=-\epsilon _{z}\sum_{i,\sigma }({\bf N}_{i,z,\sigma }-{\bf N}_{i,\bar{z}%
,\sigma }).
\]
This two parts of interaction is independent of spin.

In short, combining the JT effect and the lattice distortion caused by the
crystal field, the total effective Hamiltonian is found to be 
\begin{equation}
H_{total}=H_{e}+H_{AF}+H_{JT}+H_{z}.  \label{original}
\end{equation}

\subsection{A projection perturbation approach}

In the large $J_{H}$ limit and when the number of electrons is not greater
than the number of lattice sites, each site is occupied by at most one $%
e_{g} $ electron. And the electron must form a spin $S+1/2$ state with the
localized spin on the same site. The process can be realized by introducing
the projection operator 
\begin{equation}
P=\prod_{i}(P_{ih}+\sum_{\gamma =\pm }P_{i\gamma s}^{+}).
\end{equation}
For any state $\left| \alpha \right\rangle $, $P\left| \alpha \right\rangle $
is the component of the state with holes and single occupancies with spin $%
S+1/2$. This technique has been applied extensively to study the one-band
Kondo lattice model. \cite{Zener51,Shen97} The projection operators are
defined by 
\begin{eqnarray*}
P_{i,h} &=&\prod_{\gamma ,\sigma }(1-n_{i,\gamma ,\sigma }), \\
P_{i,s} &=&\sum_{\gamma ,\sigma }n_{i,\gamma ,\sigma }\prod_{\gamma ^{\prime
}\neq \gamma ,\sigma ^{\prime }}(1-n_{i,\gamma ^{\prime },\sigma ^{\prime
}}), \\
P_{i,\gamma ,s}^{+} &=&\sum_{\sigma ,\sigma ^{\prime }}\left( \frac{{\bf S}%
_{i}\cdot {\bf \sigma }+(S+1){\bf I}}{2S+1}\right) _{\sigma \sigma ^{\prime
}}c_{i,\gamma ,\sigma }^{\dagger }c_{i,\gamma ,\sigma ^{\prime }}P_{i,s}.
\end{eqnarray*}
Hence, in the infinite J$_{H}$ limit, the Hamiltonian in Eq. (\ref{ham}) is
reduced to 
\begin{equation}
H_{e}^{(a)}=PH_{e}P  \label{first}
\end{equation}
as shown in Fig. 1(a). This is the double exchange model with orbital
degeneracy. With the help of the Schrieffer-Wolff transformation,\cite
{Schrieffer66} the finite, but large J$_{H}$ effect is taken into account by
superexchange processes in the second-order projective perturbation
approach: (1) Mn$^{3+}$Mn$^{4+}$ $\Leftrightarrow $ Mn$^{4+}$Mn$^{3+}$as
shown in Fig. 1(b) and (2) Mn$^{3+}$Mn$^{3+}$ $\Leftrightarrow $ Mn$^{4+}$Mn$%
^{2+}$ as shown in Figs.1(c - e). After considering the second order
perturbation correction to Eq. (\ref{first}), we obtain 
\begin{equation}
H_{eff}=PHP-\sum_{\alpha }\frac{1}{\Delta E_{\alpha }}PHQ_{\alpha }HP.
\label{sec}
\end{equation}
where $\Delta E_{\alpha }$ are the energy difference of the mediate state
and the initial state as shown in Figs. 1(b-e). $Q_{\alpha }$ ($\alpha
=b,c,d,e$) are the projection operators for the intermediate states. The
derivation of Eq. (\ref{sec}) and the explicit expressions of $\Delta
E_{\alpha }$ and $Q_{\alpha }$ are presented in Appendix.

\subsection{Effective Hamiltonian in Schwinger boson representation.}

To simplify the notations, we express the Hamiltonian in the Schwinger-boson
representation.\cite{Arovas88} The representation was introduced to describe
the one-band double exchange by Sarker,\cite{Sarker96} and here we
generalize it to the our model by introducing another type of boson for
orbital degree of freedom. Define 
\[
P_{i,\gamma ,s}^{+}c_{i,\gamma ,\sigma }^{\dagger }P_{i,\gamma ,s}^{+}=\frac{%
1}{\sqrt{2S+1}}a_{i,\sigma }^{\dagger }b_{i,\gamma }^{\dagger
}f_{i}^{\dagger }
\]
where $a_{i,\sigma }^{\dagger }$ and $a_{i,\sigma }$ are the Schwinger boson
operators for spin: 
\begin{eqnarray*}
{\bf S}_{i}^{+} &=&a_{i,\uparrow }^{\dagger }a_{i,\downarrow }, \\
{\bf S}_{i}^{-} &=&a_{i,\downarrow }^{\dagger }a_{i,\uparrow } \\
{\bf S}_{i}^{z} &=&\frac{1}{2}\left( a_{i,\uparrow }^{\dagger }a_{i,\uparrow
}-a_{i,\downarrow }^{\dagger }a_{i,\downarrow }\right) .
\end{eqnarray*}
$b_{i,\alpha }^{\dagger }$ and $b_{i,\alpha }$ are the Schwinger boson
operators for orbital degrees of freedom with $\alpha =x,y,z$, which depends
on the direction of ${\bf r}_{ij}$. $b_{i,z}^{\dagger }\left| 0\right\rangle
=\left| z\right\rangle $ and $b_{i,\bar{z}}^{\dagger }\left| 0\right\rangle
=\left| \bar{z}\right\rangle .$ Other two components are not independent and
related to the z-component by a transformation: 
\[
\begin{array}{l}
b_{ix}=\frac{1}{2}b_{i,z}-\frac{\sqrt{3}}{2}b_{i,\bar{z}}, \\ 
b_{i\bar{x}}=\frac{\sqrt{3}}{2}b_{i,z}+\frac{1}{2}b_{i,\bar{z}}, \\ 
b_{iy}=\frac{1}{2}b_{i,z}+\frac{\sqrt{3}}{2}b_{i,\bar{z}}, \\ 
b_{i\bar{y}}=-\frac{\sqrt{3}}{2}b_{i,z}+\frac{1}{2}b_{i,\bar{z}}.
\end{array}
\]
$f_{i}^{\dagger }$ and $f_{i}$ are fermion operators for charge carrier: $%
n_{i}=f_{i}^{\dagger }f_{i}=1$ means that there is one charge $e$ on the
site, i.e., Mn$^{3+}$, and $n_{i}=0$ means Mn$^{4+}.$ The constraints for
the Schwinger boson and fermions are 
\begin{eqnarray*}
a_{i,\uparrow }^{\dagger }a_{i,\uparrow }+a_{i,\downarrow }^{\dagger
}a_{i,\downarrow } &=&2S+n_{i}, \\
n_{i}^{\alpha }+n_{i}^{\overline{\alpha }} &=&n_{i}.
\end{eqnarray*}
In the representation, the effective Hamiltonian for conduction electrons is
written as, 
\begin{equation}
H_{e}=H_{e}^{(a)}+H_{e}^{(b)}+H_{e}^{(c)}+H_{e}^{(d)}+H_{e}^{(e)}.
\label{total}
\end{equation}
Each term corresponds to one of the processes shown in Fig. 1, and is
expressed as 
\begin{eqnarray*}
H_{e}^{(a)} &=&-\sum_{ij,\sigma }\frac{t}{2S+1}a_{i,\sigma }^{\dagger
}a_{j,\sigma }b_{i,\alpha }^{\dagger }b_{j,\alpha }f_{i}^{\dagger }f_{j}, \\
H_{e}^{(b)} &=&\frac{2St^{2}}{J_{H}(2S+1)^{2}}\sum_{ij}\left( \frac{{\bf S}%
_{i}\cdot \widetilde{{\cal S}}_{j}-S(S+1/2)}{2S(S+1/2)}\right)
P_{ih}P_{j\alpha }, \\
H_{e}^{(c)} &=&\frac{t^{2}}{U+J_{H}S}\sum_{ij}\left( \frac{\widetilde{{\cal S%
}}_{i}\cdot \widetilde{{\cal S}}_{j}-(S+1/2)^{2}}{2(S+1/2)^{2}}\right)
P_{ij}^{s}, \\
H_{e}^{(d)} &=&\frac{t^{2}}{U^{\prime }+\frac{3J}{2}+J_{H}S}\sum_{ij}\left( 
\frac{\widetilde{{\cal S}}_{i}\cdot \widetilde{{\cal S}}_{j}-(S+1/2)^{2}}{%
2(S+1/2)(S+1)}\right) P_{ij}^{d}, \\
H_{e}^{(e)} &=&-\frac{t^{2}}{U^{\prime }-\frac{J}{2}}\sum_{ij}\left( \frac{%
\widetilde{{\cal S}}_{i}\cdot \widetilde{{\cal S}}_{j}+(S+1/2)(S+3/2)}{%
2(S+1/2)(S+1)}\right) P_{ij}^{d}
\end{eqnarray*}
where ${\bf S}_{i}$ is a spin operator with $S$, and $\widetilde{{\cal S}}%
_{i}$ is a spin operator with $S+1/2$ as a FM combination of the localized
spin and itinerant electron at the same site. The operators P are the
projection operators for charge and orbits: 
\begin{eqnarray*}
P_{ij}^{s} &=&n_{i}^{\alpha }n_{j}^{\alpha }, \\
P_{ij}^{d} &=&n_{i}^{\alpha }n_{j}^{\bar{\alpha}}, \\
P_{ih} &=&1-n_{i}, \\
P_{i\alpha } &=&n_{i}^{\alpha }
\end{eqnarray*}
where $n_{i}^{\alpha }=b_{i,\alpha }^{\dagger }b_{i,\alpha }.$ Finally, it
is worth mentioning that we just keep the two-site interactions and neglect
three-site interactions in Eq. (\ref{total}). The three-site terms describe
indirect hopping processes between the next-nearest-neighbor site via the
intermediate states and isbelieved to be relatively small compared with the
direct hopping terms.  

The other terms in Eq. (\ref{original}) become 
\begin{eqnarray*}
H_{AF} &=&J_{AF}\sum_{ij}\left( {\bf \bar{S}}_{i}\cdot {\bf \bar{S}}%
_{j}-S^{2}\right) , \\
H_{JT} &=&k\sum_{ij}(n_{i}^{\alpha }-n_{i}^{\bar{\alpha}})(n_{j}^{\alpha
}-n_{j}^{\bar{\alpha}}), \\
H_{z} &=&-\epsilon _{z}\sum_{i}(n_{i}^{z}-n_{i}^{\bar{z}}),
\end{eqnarray*}
respectively, where ${\bf \bar{S}}_{i}={\bf S}_{i}(1-n_{i})+\frac{2S}{2S+1}%
\widetilde{{\cal S}}_{i}n_{i}$.

Hence up to the order of $t^{2}/J_{H}$ the total effective Hamiltonian in
the representation of the Schwinger bosons for both spin and orbit is 
\begin{equation}
H_{eff}=H_{e}+H_{AF}+H_{JT}+H_{z}.  \label{eff}
\end{equation}
Here each term should be restricted in the projected space. Approximately, $%
H_{eff}$ in Eq. (\ref{eff}) and $H_{total}$ in Eq. (\ref{original}) are
expected to describe the same low energy physics in the large J$_{H}$ case.

The present theory is based on Eq. (\ref{eff}), in terms of which we are
able to establish a unified description for the electronic behaviors in
doped manganites for the first time. In this paper, the model parameters are
roughly estimated from the excitation energies of Mn ions and the functional
density calculations:\cite{Griffith71} we take $t=0.41eV$ as energy unit. $%
2t/J_{H}(2S+1)=0.35$; $t/(U+J_{H}S)=0.042$; $t/(U^{\prime }+\frac{3}{2}%
J+J_{H}S)=0.056$; $t/(U^{\prime }-\frac{1}{2}J)=0.106$; and $J_{AF}=0.001$.
All the phase diagrams in this paper are established on this set of
parameters.

\section{Spin-Charge-Orbital Coherent State Formalism}

The second order projection perturbation approach includes part of strong
correlation between conduction electrons and localized spin, and removes the
direct Hund coupling J$_{H}$. Some properties of the model [Eq. (\ref{eff})]
become clearer than the original one [Eq. (\ref{ham})]. For instance, H$%
_{e}^{(a)}$ describes the double exchange mechanism for ferromagnetism; H$%
_{e}^{(b)}$ describes a particle-hole interaction with an AF coupling; H$%
_{e}^{(c)}$ and H$_{e}^{(d)}$describe the AF superexchange couplings with
the same and different orbitals, respectively, and H$_{e}^{(e)}$ describes a
FM\ superexchange coupling with different orbitals. Each term becomes
predominate at some point in the phase space. However, the effective
Hamiltonian [Eq. (\ref{eff})] still seems to be very complicated, and it is
still very hard to fix its physical properties. From the study of the famous
t-J model, which is derived from the one-band Hubbard model by the same
projection perturbation, we learn that this is just a first step to
understand physics of the electronic Hamiltonian in Eq. (\ref{ham}) in the
strong Hund coupling case.

\subsection{Spin-charge-orbital coherent state}

To investigate the effective Hamiltonian, we apply the spin-coherent state
mean-field theory. We introduce two polar parameters $\theta _{i}$ and $\phi
_{i}$ for spin bosons at the site $i$, and two parameters $\alpha _{i}$ and $%
\beta _{i}$ for orbital bosons. Following Auerbach,\cite{Auerbach94} we
define the spin-charge-orbital coherent state 
\begin{eqnarray*}
&&\left| \theta _{i},\phi _{i},\alpha _{i},\beta _{i},\xi _{i}\right\rangle
=\left| \theta _{i},\phi _{i}\right\rangle _{S}\left| 0\right\rangle \\
&&+\left| \theta _{i},\phi _{i}\right\rangle _{S+1/2}\otimes \left| \alpha
_{i},\beta _{i}\right\rangle _{1/2}\xi _{i}f_{i}^{\dagger }\left|
0\right\rangle
\end{eqnarray*}
where 
\begin{eqnarray*}
&&\left| \theta _{i},\phi _{i}\right\rangle _{S}=\frac{1}{\sqrt{(2S)!}} \\
&&\times \left[ \cos \frac{\theta _{i}}{2}e^{i\phi _{i}/2}a_{i,\uparrow
}^{\dagger }+\sin \frac{\theta _{i}}{2}e^{-i\phi _{i}/2}a_{i,\downarrow
}^{\dagger }\right] ^{2S}\left| 0\right\rangle , \\
&&\left| \alpha _{i},\beta _{i}\right\rangle _{1/2}=\left( \cos \frac{\alpha
_{i}}{2}e^{i\beta _{i}/2}b_{i,z}^{\dagger }+\sin \frac{\alpha _{i}}{2}%
e^{-i\beta _{i}/2}b_{i,\bar{z}}^{\dagger }\right) \left| 0\right\rangle
\end{eqnarray*}
and $\xi $ is an anticommuting Grassmann variable. Define 
\[
\left| \Phi \right\rangle =\prod \bigotimes \left| \theta _{i},\phi
_{i},\alpha _{i},\beta _{i},\xi _{i}\right\rangle . 
\]
The Hamiltonian function is 
\[
{\cal H}=\frac{\left\langle \Phi \right| H_{eff}\left| \Phi \right\rangle }{%
\left\langle \Phi \right| \left. \Phi \right\rangle } 
\]
where

\begin{eqnarray*}
{\cal H}_{e}^{(a)} &=&-\sum_{ij}t(t_{ij}^{s}t_{ij}^{\alpha }\xi _{i}^{\ast
}\xi _{j}+h.c.), \\
{\cal H}_{e}^{(b)} &=&-\frac{2St^{2}}{J_{H}(2S+1)^{2}}\sum_{ij}\sin ^{2}%
\frac{\Theta _{ij}}{2}n_{j,\alpha }\left( 1-\xi _{i}^{\ast }\xi _{i}\right)
\xi _{j}^{\ast }\xi _{j}, \\
{\cal H}_{e}^{(c)} &=&-\frac{t^{2}}{U+J_{H}S}\sum_{ij}\sin ^{2}\frac{\Theta
_{ij}}{2}n_{i,\alpha }n_{j,\alpha }\xi _{i}^{\ast }\xi _{i}\xi _{j}^{\ast
}\xi _{j}, \\
{\cal H}_{e}^{(d)} &=&-\frac{t^{2}}{U^{\prime }+\frac{3J}{2}+J_{H}S}\frac{%
(S+1/2)}{(S+1)} \\
&&\times \sum_{ij}\sin ^{2}\frac{\Theta _{ij}}{2}n_{i,\alpha }n_{j,\bar{%
\alpha}}\xi _{i}^{\ast }\xi _{i}\xi _{j}^{\ast }\xi _{j}, \\
{\cal H}_{e}^{(e)} &=&-\frac{t^{2}}{U^{\prime }-\frac{J}{2}}\sum_{ij}\left( 
\frac{(S+1/2)\cos \Theta _{ij}+(S+3/2)}{2\left( S+1\right) }\right) \\
&&\times n_{i,\alpha }n_{j,\bar{\alpha}}\xi _{i}^{\ast }\xi _{i}\xi
_{j}^{\ast }\xi _{j}, \\
{\cal H}_{AF} &=&-2J_{AF}S^{2}\sum_{ij}\sin ^{2}\frac{\Theta _{ij}}{2}, \\
{\cal H}_{JT} &=&k\sum_{ij}(n_{i,\alpha }-n_{i,\bar{\alpha}})(n_{j,\alpha
}-n_{j,\bar{\alpha}})\xi _{i}^{\ast }\xi _{i}\xi _{j}^{\ast }\xi _{j}, \\
{\cal H}_{z} &=&-\epsilon _{z}\sum_{i}(n_{i,z}-n_{i,\bar{z}})\xi _{i}^{\ast
}\xi _{i}
\end{eqnarray*}
where 
\begin{eqnarray*}
t_{ij}^{s} &=&\cos \frac{\theta _{i}}{2}\cos \frac{\theta _{j}}{2}e^{i\frac{%
\phi _{i}-\phi _{j}}{2}}+\sin \frac{\theta _{i}}{2}\sin \frac{\theta _{j}}{2}%
e^{-i\frac{\phi _{i}-\phi _{j}}{2}}, \\
t_{ij}^{x} &=&\left( \frac{1}{2}\cos \frac{\alpha _{j}}{2}e^{-i\beta _{j}/2}-%
\frac{\sqrt{3}}{2}\sin \frac{\alpha _{j}}{2}e^{i\beta _{j}/2}\right) \\
&&\times \left( \frac{1}{2}\cos \frac{\alpha _{i}}{2}e^{i\beta _{i}/2}-\frac{%
\sqrt{3}}{2}\sin \frac{\alpha _{i}}{2}e^{-i\beta _{i}/2}\right) , \\
t_{ij}^{y} &=&\left( \frac{1}{2}\cos \frac{\alpha _{j}}{2}e^{-i\beta _{j}/2}+%
\frac{\sqrt{3}}{2}\sin \frac{\alpha _{j}}{2}e^{i\beta _{j}/2}\right) \\
&&\times \left( \frac{1}{2}\cos \frac{\alpha _{i}}{2}e^{i\beta _{i}/2}+\frac{%
\sqrt{3}}{2}\sin \frac{\alpha _{i}}{2}e^{-i\beta _{i}/2}\right) , \\
t_{ij}^{z} &=&\cos \frac{\alpha _{i}}{2}\cos \frac{\alpha _{j}}{2}e^{i(\beta
_{i}-\beta _{j})/2}, \\
n_{i,\alpha } &=&\cos ^{2}\left( \frac{\alpha _{i}}{2}+\delta _{\alpha
}\right) -\sin 2\delta _{\alpha }\sin \alpha _{i}\sin ^{2}\frac{\beta _{i}}{2%
}, \\
n_{i,\bar{\alpha}} &=&\sin ^{2}\left( \frac{\alpha _{i}}{2}+\delta _{\alpha
}\right) +\sin 2\delta _{\alpha }\sin \alpha _{i}\sin ^{2}\frac{\beta _{i}}{2%
}
\end{eqnarray*}
with 
\[
\cos \Theta _{ij}=\cos \theta _{i}\cos \theta _{j}+\sin \theta _{i}\sin
\theta _{j}\cos (\phi _{i}-\phi _{j}) 
\]
and $\delta _{x}=-\pi /3$, $\delta _{y}=\pi /3$, and $\delta _{z}=0.$

\subsection{Mean Field Approximation}

${\cal H}$ includes the fourth powers of Grassmann variables. These terms
are hard to integrate, and an approximation is needed. 
\[
f_{i}^{\dagger }f_{i}f_{j}^{\dagger }f_{j}\approx \left\langle
f_{i}^{\dagger }f_{i}\right\rangle f_{j}^{\dagger }f_{j}+f_{i}^{\dagger
}f_{i}\left\langle f_{j}^{\dagger }f_{j}\right\rangle -\left\langle
f_{i}^{\dagger }f_{i}\right\rangle \left\langle f_{j}^{\dagger
}f_{j}\right\rangle . 
\]
$\left\langle f_{i}^{\dagger }f_{i}\right\rangle $ is taken to the density
of electrons in a density-uniform state, and of sublattice-dependence when
we consider a charge ordering state. The polar parameters are also treated
in the mean field theory, following de Gennes.\cite{Gennes60} The angles
between two nearest-neighbor spins are taken to be $(-1)^{i({\bf r}_{i}-{\bf %
r}_{j}){\bf \pi }}Q_{\alpha }.$ If all $Q_{\alpha }$ are equal to zero, it
is a FM state. For a A-type AF, $Q_{\alpha }$ are zero among the x-y plane,
and $Q_{z}$ is non-zero along the z-axis. It is worth stressing that in this
paper the antiferromagnetism along a specific direction does not mean that $%
Q_{\alpha }$ must be $\pi $. For a C-type AF, $Q_{z}$ is taken to be zero
and $Q_{x,y}$ are non-zero. G-type AF means that all $Q_{\alpha }$ are
equal, but non-zero. The parameters for orbital degrees of freedom are
treated similarly as those for spins. In this paper, we limit our discussion
only in the case of $T=0$, i.e., the ground state. The phase diagrams are
established by minimizing the free energy.

\section{Charge Ordering and Magnetic Structures}

Using the spin-charge-orbital coherent state theory, we focus on three
phenomena which occur at different densities of doping: (a) ferromagnetism
and A-type antiferromagnetism at $x=0$; (b) phase separation at small
doping; and (c) charge ordering at $x=0.5$.

\subsection{Ferromagnetism at $x=0$}

The magnetic structure of the parent compound ($x=0$) has also very decisive
impact on the electronic properties near the point, and is sensitive to the
model parameters. The effective spin superexchange coupling is approximately 
\begin{eqnarray*}
&&J_{eff}/t=\frac{t^{2}}{U+J_{H}S}\frac{\left\langle P_{ij}^{s}\right\rangle 
}{(S+1/2)^{2}}+\frac{S^{2}}{\left( S+1/2\right) ^{2}}J_{AF} \\
&&-\left( \frac{t^{2}}{U^{\prime }-\frac{J}{2}}-\frac{t^{2}}{U^{\prime }+%
\frac{3J}{2}+J_{H}S}\right) \frac{\left\langle P_{ij}^{d}\right\rangle }{%
2\left( S+1/2\right) \left( S+1\right) },
\end{eqnarray*}
which depends not only on the model parameters, but also the orbital
orderings. It is worth noticing that the factor before $\left\langle
P_{ij}^{d}\right\rangle $ is always negative, and the factor before $%
\left\langle P_{ij}^{s}\right\rangle $ is positive. When U is taken to be
infinite, the factor before $\left\langle P_{ij}^{s}\right\rangle $
vanishes. In other words, the AF coupling in H$_{e}^{(c)}$ is suppressed
completely. As the FM\ coupling in H$_{e}^{(e)}$ is always stronger than the
AF coupling in H$_{e}^{(d)}$ (i.e., the factor before $\left\langle
P_{ij}^{d}\right\rangle $ is always negative), the FM coupling becomes
predominant, and the ground state becomes FM at low temperatures if we do
not take into account the tiny AF coupling of $J_{AF}$.\cite{Shen99} In the
case, the magnetic structure is independent of the JT effect and the lattice
distortion, {\it i.e.}, the orbital distribution of conduction electrons.
When U is finite, the strength $k$ of the JT effect and the lattice
distortion coefficient $\epsilon _{z}$ affect the magnetic structure by
adjusting orbital ordering: the JT effect favors to form a ``Neel's type
AF'' orbital-orbital correlation, which enhances the FM\ coupling through
the process (e) in Fig. 1(e), while the lattice distortion tends to force
the orbital to polarize, which increases the AF coupling through the
processes ( b - d ) in Fig. 1(b-d). The phase diagram in Fig. 2 shows that
the FM\ survives at finite U for a small $\epsilon _{z}$ and large $k$ and
evolves into A-type AF for a large $\epsilon _{z}$ and small $k$ (We choose
the model parameters in Section III.E.) Hence the ferromagnetism at $x=0$
comes from the superexchange process in Fig. 1 (e), not the double exchange
mechanism in Fig. 1 (a) as the direct hopping is prohibited due to the
strong coupling at $x=0$. The anisotropy of magnetism at x=0 originates from
the crystal field along the z-direction, which forces orbital degrees to
form a ``FM''-like state along that direction while the ``AF'' ordering
remains among the z-y planes. It is worth mentioning that the ferromagnetic
superexchange mechanism appears only when the orbital degeneracy of e$_{g}$
electrons is taken into account. For a simplified one-band ferromagnetic
Kondo lattice model, the ground state at half filling is always
antiferromagnetic for any finite J$_{H}$ and U.\cite{Shen96} Hence to
understand the ferromagnetism, we have to take the orbital degeneracy of e$%
_{g}$ electrons into account such that the ferromagnetic superexchange
coupling, which originates from the virtue process in Fig. 1(e), occurs,
meanwhile the double exchange ferromagnetism is compressed completely.\cite
{Slater36}

\begin{figure}[tbp]
\epsfxsize=8.5cm
\epsfbox{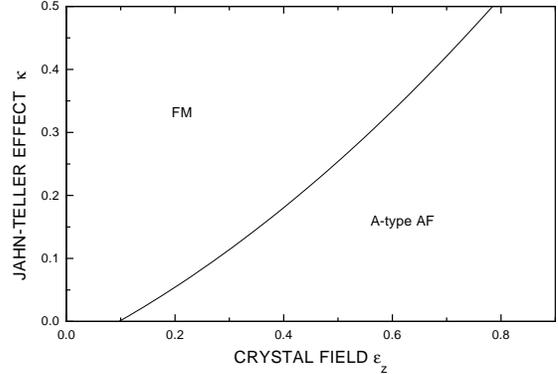}
\caption{ The phase diagram at $x=0$.}
\end{figure}

\subsection{Phase separation and A-type antiferromagnetism at $x\rightarrow
0 $}

It was reported experimentally that the phase separation was observed in the
single crystal of La$_{1-x}$Ca$_{x}$MnO$_{3}$ at $x=0.05$ and $0.08$.\cite
{Allodi97} In the present theory, FM coupling is very strong near $x=0$ as
shown Fig. 2 . Under the ferromagnetic background, the AF couplings in H$_{e}
$ and H$_{AF}$ are suppressed. The only interacting term survived for
polarized particles is H$_{e}^{(e)}$ in Eq. (\ref{eff}). Considering that
the term vanishes unless the neighboring sites are occupied by two particles
on different orbits at the same time, a pure attraction arises between the
charge carriers. When the lattice distortion increases, FM\ phase evolves
into an A-type AF as the distortion forces the orbital boson to polarize
along the c-axis, which further enhances the AF coupling via the process
(b). In the case of A-type AF, it is FM within the x-y plane, and these FM
planes are coupled antiferromagnetically. As there is no hopping between two
layers with opposite spins, the system can be regarded as a reduced
2-dimensional one. Phase separation was discussed in the one-band Kondo
lattice model numerically and analytically \cite{Yunoki97,Shen98,Arovas98}
and is associated with the AF structure. To explore the physical origin of
PS near $x=0$, the orbital degeneracy of the e$_{g}$ electrons has to be
taken into account as the PS arises under the FM background. The FM near $%
x\rightarrow 0$ in x-y plane originates from H$_{e}^{(e)}$ in Eq. (\ref{eff}%
), {\it i.e.}, the spin FM superexchange process. When the system deviates
from the point slightly the magnetic structure should not change
qualitatively. In fact, the PS was indeed observed in an A-type AF\
background.\cite{Allodi97} When the FM forms in the x-y plane, the
interactions for charge carriers related to AF coupling are suppressed, and
the attractive interaction in the H$_{e}^{(e)}$ becomes predominant. On the
other hand, the orbital bosons tend to form an orbital ``AF'' state, which
further suppresses the effective hopping term. This property will enhance
the relative ratio of the attraction to the effective hopping. The strong
attraction \ is the physical origin to the phase separation: the charge
carriers will evolve into two regimes with high and low density of the
carriers. In the case, the regime with the high density of charge carriers
has a FM background, and the regime with the low density has an AF
background as only the J$_{AF}$ term survives at $x=0$. Along the c-axis,
the antiferromagnetic structure will suppress H$_{e}^{(a)}$ and H$_{e}^{(e)}$%
. Thus the pure interaction along the c-axis is repulsive. In reality, the
angle between the spins on different layer is not absolutely $\pi $. So the
dimensionality of A-type AF\ should\ be between 2 and 3. It is known that a
higher dimensionality favors to form PS.\cite{Emery90} The canted FM along
the c-axis would enhance the stability of PS. The phase diagram of PS with
respect to the JT effect and lattice distortion is shown in Fig. 3. When the
lattice distortion increases, a stronger JT effect is required to induce PS,
as expected from our general discussion. Yunoki et al.\cite{Yunoki98}
recently reported that the PS appears in the phase diagram of a
two-dimensional model with the Monte-Carlo method. Our result confirms their
numerical prediction. However, the on-site Coulomb interaction was not
included in their calculation.

\begin{figure}[tbp]
\epsfxsize=8.5cm
\epsfbox{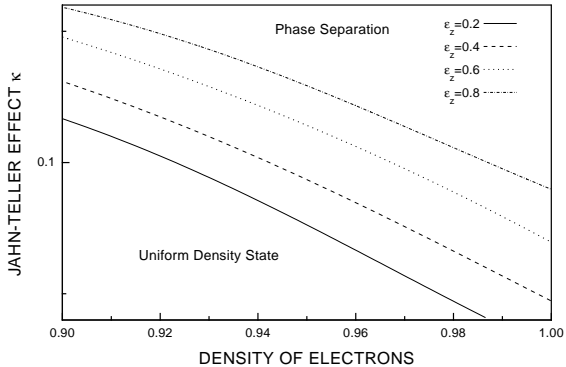}
\caption{Phase separation and the Jahn-Teller effect near $x=0$.}
\end{figure}

\subsection{Charge ordering and antiferromagnetism at $x=0.5$}

Occurrence of the charge ordering at $x=0.5$ is associated with AF
structure. For instance, the charge ordering with the $(\pi ,\pi ,0)$
pattern occurs under the background of C-type AF, that is, AF in the x-y
plane and FM along the c-axis.\cite{Jirak85} The main feature of the charge
ordering with $(\pi ,\pi ,0)$ is that Mn$^{4+}$ and Mn$^{3+}$ ions aligns
regularly in the x-y plane and the Mn$^{3+}$ ions align along the c-axis. It
is well known that the CO state is expected to be stabilized when the
repulsive interaction between charge carriers dominates the kinetic energy.
The particle-hole interaction \ in the process (b) [or H$_{e}^{(b)}$ in Eq. (%
\ref{eff}) ] is approximately proportional to $x(1-x)$ and reaches a minimal
value at $x=0.5$. As the sign of the particle-hole interaction is negative,
it is equivalent to a repulsive interaction between charge carriers (or
holes). We put forward that the physical origin of the charge ordering
results from this process. Of course the direct nearest neighbor Coulomb
interaction is also favorable for the charge carriers to form CO state, as
some authors discussed.\cite{Tomioka95,Lee97} The phase diagram for charge
ordering at $x=0.5$ shown in Fig. 4 depends on the JT effect and lattice
distortion as well as the parameters for electronic interactions. In the
case of C-type AF, the particle-hole interaction in H$_{e}^{(b)}$ becomes
stronger, and the hopping term is also suppressed as the e$_{g}$ electron
cannot hop to a site with antiparallel spin. Relatively, the effective
interaction becomes divergent when the spin-spin correlation becomes AF. In
this case the state with a uniform density is unstable against the CO. To
minimize the potential energy, the charge carriers tend to form the CO
within the x-y plane. Along the c-axis, the FM structure makes the effective
interaction attractive and all charge carriers will accumulate along the
axis. Therefore, the CO has the $(\pi ,\pi ,0)$ pattern. When the JT effect
becomes weaker and the lattice distortion increases, the AF coupling
increases such that the FM coupling along the C-axis is suppressed. In the
case the effective interaction along the c-axis also becomes repulsive.
Therefore, the G-type AF with the $(\pi ,\pi ,\pi )$ pattern should be
stable. A stronger JT effect enhances FM coupling while a stronger lattice
distortion increases AF coupling along the c-axis. Thus it will force the
C-type AF to evolve to the A-type AF. The charge carriers have the $(0,0,\pi
)$ pattern as the interaction is repulsive for AF coupling and attractive
for FM coupling.

\begin{figure}[tbp]
\epsfxsize=8.5cm
\epsfbox{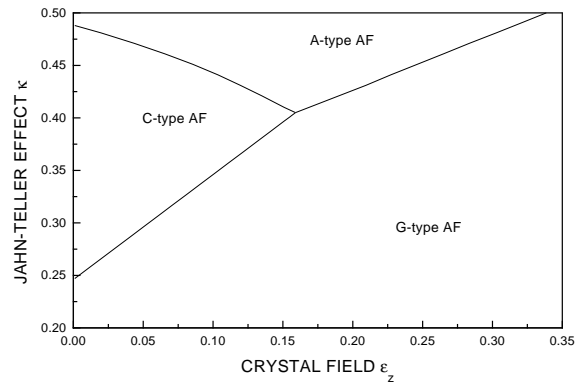}
\caption{ The stable magnetic structure at $x=0.5$. The G-type AF has a
charge carriers distribution with the pattern ($\protect\pi $, $\protect\pi $%
, $\protect\pi $). The C-type AF has the pattern ($\protect\pi $, $\protect%
\pi $, 0). And the A-type AF has a layered pattern (0, 0, $\protect\pi $) }
\end{figure}

\section{Discussion and Summary}

Here we discuss the relation between the on-site U and sign of effective
interactions. The phase diagram in this paper bases on the parameters of the
model which we list at the end of Section II. The parameters are roughly
estimated from the excitation energies of Mn ions and the functional density
calculations. However the model will contain richer phases if we adjust the
model parameters.\ Let us first see two limits. (a) $U$, $U\prime \gg J_{H}S$%
, $J$. From the excitation energies of intermediate states we list in Table
I, the energies in the states of Fig. 1(c-d) will be much higher than that
of the state in Fig. 1(b). The FM superexchange coupling will be suppressed,
and only the AF particle-hole superexchange coupling survives. In this case,
it favors forming CO in the vicinity of $x=0.5$, but does not favor driving
charge carriers to phase separate near $x=0$ as , equivalently, the net
particle-particle interaction is repulsive. (b) $U\ll J_{H}S$. The
intermediate state in Fig. 1(b) will have a higher energy, and the process
will be suppressed. The main competition comes from the process (c) and (d).
The former is AF, and the later is FM. Both of them induce an attractive
interaction for charge carriers. In this case, it favors forming PS near $%
x=0 $, but does not favor forming CO at$\ x=0.5.$ The density of dopants of
divalent ions X in R$_{1-x}$X$_{x}$MnO$_{3}$ will induce structural
parameters of the crystal, but should have little impact on the model
parameters related interactions in the ions. Hence a mediate value of
on-site interaction is very important to explain PS and CO simultaneously in
the same model with the same model parameters. The parameters we used in
this paper is in the mid-way of the two limits. The ratio of U to J$_{H}$S
is roughly estimated to be 4.8, which indicates that the on-site interaction
U will also have an important impact in the formation of the rich phases in
the doped manganites. Detailed discussion on the model parameters for
transition-metal oxides is referred to Ref. \cite{Imada98}

Next, for doped manganites, the on-site interaction is so strong that the
system is one of typical strongly correlated electron systems. In our
second-order projection perturbation approach, part of electron correlations
has been taken into account, and the direct on-site Coulomb interaction and
Hund's rule coupling are removed. The physical meanings in each term of the
effective Hamiltonian in Eq. (\ref{eff}) become much clearer than that in
Eq. (\ref{original}). We can see clearly the physical origins of various
type magnetic structures and related physical processes. Strictly speaking,
the effective Hamiltonian in Eq. (\ref{eff}) and the original Hamiltonian in
Eq. (\ref{original}) are equivalent ONLY in the limit of large $U$ and $%
J_{H}S$. Up to the order of t/U and $t/J_{H}S$, we expect that the two
Hamiltonians describe the same physics at low temperatures based on the
principle of the perturbation theory. On the other hand, although the
projection technique is proved to be one of powerful tools to deal with
strongly correlated electron systems,\cite{Fulde95} it is still a
challenging problem to deal with the effective Hamiltonian. Our
spin-orbital-charge coherent state theory is just an initial step to
understand the physics in doped manganites.

Finally, we come to discuss the relation of the charge inhomogeneity and the
long-range Coulomb interaction. The tendency to PS and CO depends on the
magnetic structures. It is anticipated that the inclusion of the
longer-range Coulomb interaction will lead to a stable and microscopically
inhomogeneous state.\cite{Moreo99} Recently Mori {\it et al.} reported that
charge stripes arise in the range of $x>0.5$, and tend to form stripe pairs. 
\cite{Mori98} It reveals strong repulsion for the charge carriers since the
striped Mn$^{3+}$ ions are separated by Mn$^{4+}$. By considering the
nearest neighbor interaction and longer-range Coulomb interaction, L\"{o}w 
{\it et al.} obtained some numerical evidences to support the stripe phase
at x=0.5 in a simplified lattice model.\cite{Low94} Our electronic model
does not include the long-range Coulomb interaction. However even if we take
the interaction into account, the nearest neighbor interaction should not be
very strong such that it does not destroy the tendency to PS in the vicinity
of $x=0$ caused by the superexchange attraction. Of course, it enhances the
tendency to CO near $x=0.5$ as some authors argued.\cite{Tomioka95,Lee97}

In short, starting from a realistic model, we derive an effective
Hamiltonian by means of the second order projection perturbation approach in
the case of the strong Hund coupling. In order to treat the model, we
introduce a new type of bosons for orbital degree of freedom as well as
bosons for spins and fermions for charge carriers. A spin-charge-orbital
coherent state theory is presented. Physically, by adjusting the orbital
ordering of the charge carriers, we find that the JT effect and the lattice
distortion have strong impact on the electronic collective behaviors as well
as the magnetic structures. At the undoped case ($x=0$), the ferromagnetism
originates from the FM superexchange coupling, and the anisotropy of A-type
AF is induced by the crystal field. Away from the undoped case, the FM
superexchange coupling term is responsible to form PS near the slight doping
regime while the AF particle-hole interaction drives the charge carriers to
form CO near $x=0.5$.

This work was supported by a CRCG research grant at the University of Hong
Kong.

\begin{appendix}

\section{The Second-Order Perturbation Expansion}

In this Appendix, we derive the effective Hamiltonian in Eq.(\ref{sec}) in
which correction of the finite but large $J_{H}$ effect is taken into
account. We follow Schrieffer and Wolff's method \cite{Schrieffer66} to
derive the Kondo Hamiltonian from the periodic Anderson Hamiltonian (Also
see Ref.\cite{Fulde95}). According to the projection operator $P$, the
Hilbert space is divided into a subspace $P$, which consists of holes and
single occupancies with spin $S+1/2$, \ and a subspace $Q(=1-P)$ with at
least one double occupancy or one single occupancy with $S-1/2$. The Schr%
\"{o}dinger equation for the system is written as 
\[
H\left| \phi \right\rangle =E_{g}\left| \phi \right\rangle 
\]
where $E_{g}$ is the ground state energy. The equation can be expressed in
the two subspaces $P$ and $Q$%
\begin{eqnarray}
PHP\left| \phi \right\rangle +PHQ\left| \phi \right\rangle  &=&E_{g}P\left|
\phi \right\rangle ,  \label{pq-1} \\
QHP\left| \phi \right\rangle +QHQ\left| \phi \right\rangle  &=&E_{g}Q\left|
\phi \right\rangle .
\end{eqnarray}
The Hamiltonians $PHP$ and $QHQ$ act within the subspaces $P$ and $Q$,
respectively. $PHQ$ and $QHP$ connect the two subspaces. To eliminate the
state $Q\left| \phi \right\rangle $ in Eq.(\ref{pq-1}), we reduce the
problem in the subspace $P$.

\[
(H_{P}-E_{g})P\left| \phi \right\rangle =0 
\]
where 
\[
H_{P}=PHP-PHQ\frac{1}{QHQ-E_{g}}QHP. 
\]
The operator $Q$ can be expanded as 
\begin{eqnarray*}
Q &=&\sum_{i,\alpha }Q_{i,\alpha }\left\{ \prod_{j\neq
i}(P_{jh}+\sum_{\gamma =\pm 1}P_{j,\gamma ,s}^{+})\right. \\
&&\left. +\sum_{j,\alpha ^{\prime }}Q_{j,\alpha ^{\prime }}\prod_{k\neq
i,j}(P_{kh}+\sum_{\gamma =\pm 1}P_{k,\gamma ,s}^{+})+\cdots \right\} \\
&=&\sum_{\alpha }Q_{\alpha }.
\end{eqnarray*}
where $Q_{\alpha }$ is the projection operator that there is at least one
double occupancy or single occupancy with spin $2S-1.$ For our purpose, we
just consider the energy correction up to the second order perturbation.
Hence we take approximately 
\[
Q_{\alpha }=\sum_{i}Q_{i,\alpha }\prod_{j\neq i}(P_{jh}+\sum_{\gamma =\pm
1}P_{j,\gamma ,s}^{+}). 
\]
One of important properties is 
\[
Q_{i,\alpha }Q_{i,\alpha ^{\prime }}=Q_{i,\alpha }\delta _{\alpha \alpha
^{\prime }}. 
\]

As we are merely concerned with the low-energy excitation, the term 
\[
Q_{\alpha }\frac{1}{QHQ-E_{g}}Q_{\alpha } 
\]
is replaced approximately by 
\[
\frac{1}{\Delta E_{\alpha }}Q_{\alpha } 
\]
where $\Delta E_{\alpha }$ is the energy difference of energy with one $%
Q_{\alpha }$ and energy of $PHP$. Thus the Hamiltonian is reduced to 
\[
H_{eff}=PHP-\sum_{\alpha }\frac{1}{\Delta E_{\alpha }}PHQ_{\alpha }HP. 
\]
In the large U one-band Hubbard model, we have only one projection operator
for the intermediate state,{\it \ i.e}., the operator for the double
occupancy. In our case, we consider four intermediate states. The four
energy differences and projection operator $Q_{i,\alpha }$ of the
intermediate state shown in Fig.1 (b-e) are listed in the Table I.

\begin{table}[tbp]
\caption{The four energy differences and projection operator $Q_{i,\protect%
\alpha }$ of the intermediate state shown in Fig. 1 (b-e).}
\begin{tabular}{||c|c|c||}
$\alpha $ & Energy $\Delta E_{\alpha }$ & Operator $Q_{i,\alpha }$ \\ \hline
b & $J_{H}(2S+1)$ & $\left( \frac{-{\bf S}_{i}\cdot {\bf \sigma }+S{\bf I}}{%
2S+1}\right) _{\sigma \sigma ^{\prime }}c_{i,\gamma ,\sigma }^{\dagger
}c_{i,\gamma ,\sigma ^{\prime }}P_{i,s}$ \\ \hline
c & $U+J_{H}S$ & $n_{i,\gamma ,\uparrow }n_{i,\gamma ,\downarrow }$ \\ \hline
e & $U^{\prime }+\frac{3}{2}J+J_{H}S$ & $\left( \frac{-{\bf \tilde{S}}%
_{i}\cdot {\bf \sigma }+\left( S+1/2\right) {\bf I}}{2S+2}\right) _{\sigma
\sigma ^{\prime }}c_{i,\gamma ,\sigma }^{\dagger }c_{i,\gamma ,\sigma
^{\prime }}P_{i,\bar{\gamma},s}$ \\ \hline
d & $U^{\prime }-\frac{1}{2}J$ & $\left( \frac{{\bf \tilde{S}}_{i}\cdot {\bf %
\sigma }+\left( S+3/2\right) {\bf I}}{2S+2}\right) _{\sigma \sigma ^{\prime
}}c_{i,\gamma ,\sigma }^{\dagger }c_{i,\gamma ,\sigma ^{\prime }}P_{i,\bar{%
\gamma},s}$%
\end{tabular}
\end{table}

\end{appendix}

\end{document}